\documentstyle[11pt,epsf]{article}
\input{psfig}
\newcommand{\ds}{\displaystyle}
\begin{document}
\title{\bf{Sandpiles on a Sierpinski gasket}}
\author{Frank Daerden, Carlo Vanderzande\\ \\Departement Wiskunde 
Natuurkunde Informatica\\Limburgs Universitair Centrum\\3590 
Diepenbeek,Belgium}
\maketitle
\ \\
\begin{abstract}
We perform extensive simulations of the sandpile model on a Sierpinski gasket.
Critical exponents for waves and avalanches are determined. We extend
the existing theory of waves to the present case. This leads to an
exact value for the exponent $\tau_w$ which describes the distribution of
wave sizes: $\tau_w = \ln{(9/5)}/\ln{3}$. Numerically, it is
found that the number of waves in an avalanche is proportional to
the number of distinct sites toppled in the avalanche. This leads to a conjecture
for the exponent $\tau$ that determines the distribution of
avalanche sizes: $\tau=1+\tau_w = \ln{(27/5)}/\ln{3}$. Our predictions are in good agreement
with the numerical results.
\end{abstract}
\newpage
\ \\
PACS-classification: 64.60.Ak, 5.40.+j, 5.70.Ln
\ \\
\ \\
corresponding author:\\
Carlo Vanderzande\\
Departement WNI\\
Limburgs Universitair Centrum\\
Universitaire Campus\\
3590 Diepenbeek\\
Belgium\\
\ \\
telephone:32/11/268216\\
fax:32/11/268299\\
\ \\
e-mail:cvdzande at luc.ac.be
\newpage
\section{Introduction}
The sandpile model \cite{Bak1} plays a special role in non-equilibrium statistical
mechanics since it is the standard example of a system that reaches
a critical stationary state without the necessity of fine-tuning any parameters.
This property, which was named self organised criticality, has been
suggested to lie at the basis of the widespread occurence of fractals and $1/f$-noise
in natural systems \cite{Bak2}. Though the sandpile model seems to have little relevance
for the description of real sand, it has the advantage that it can be
threated with analytical methods. Hence, it plays a role similar to that
of the Ising model in the theory of equilibrium critical phenomena.

In the ten years that have passed since its introduction the model has been
intensively studied, mostly in two dimensions. Besides extensive numerical
work \cite{Manna}, several theoretical developments have been made [4--8]. These have shown that 
many properties of the model are closely connected to those of spanning trees,
which in turn can be described by the ($q \to 0$)--limit of the Potts model \cite{DharM}.
These connections have, for the case of two dimensions, led to exact results for the exponents
that describe the distribution of sizes of waves and last waves \cite{LastW,Waves}.
Moreover, the relation to spanning trees has led Priezzhev and his coworkers \cite{Priez} to a conjecture (again in $d=2$) for the
exponent governing for the sizes of avalanches, which
in a way is the `most wanted' of the sandpile exponents. The conjectured value
is in good agreement with numerical estimates.
In the present paper, we present results for the sandpile model on the Sierpinski
gasket, a well known deterministic fractal. We have two main reasons for doing this.
Firstly, we want to extend the theory for waves and the conjecture for
avalanches to the case of the Sierpinski gasket and compare them with the
results coming from extensive simulations. This can also lead
to a better appreciation of the results in two dimensions.
Secondly, the development of effective
renormalisation group approaches is a major goal in the field of non-equilibrium statistical
mechanics. In this respect deterministic fractals can play an important role since
often real space renormalisation approaches can be worked out exactly for them.
Some real space approaches to sandpile models have been introduced and the
results presented here can be useful in testing their validity \cite{Piet,Wies}.
We also mention that we are not the first to study the sandpile model
on a Sierpinski gasket \cite{Zap}. A previous study was however purely numerical and
was moreover restricted to an investigation of avalanches.

This paper is organised as followed. In section II we introduce the sandpile model,
and give a brief overview of the existing theoretical work. In section III, we
present our numerical results for waves and avalanches on the Sierpinski gasket.
In section IV, our data are compared with exact results which we can obtain for wave
exponents. Based on our numerical results, we also give a conjecture for the exponent $\tau$.
Finally we present some concluding remarks in section V.

\section{Sandpiles, spanning trees and waves}
The (Abelian) sandpile model can be defined on any graph, but for definiteness
we will introduce it in the contexts of the square lattice and of the Sierpinski
gasket. In both cases each vertex of the lattice has four nearest neighbours.
To each such vertex $i$ we associate a height variable $z_i$ which can take on
any positive integer number. We also introduce a critical height $z_c$ of the
height variable which we will take equal to four for all vertices.
Furthermore, it is important to stress that we work on a finite part
of the lattice (which we will denote by ${\cal L}$), which contains $N$ vertices.
In the case of the Sierpinski gasket, $N$ is trivially related to the number of iterations
$n$ used in constructing the fractal.

The sandpile model has two time scales. On a
slow time scale we drop sand at a randomly selected site and thereby increase the height
variable by one unit: $z_i \to z_i +1$. When at a given site, $z_i > z_c$,
we say that the site becomes {\it unstable} and perform a {\it toppling} 
of that site which involves the following operations:
\begin{eqnarray}
z_i & \to & z_i + \Delta_{i,j}
\label{2.1}
\end{eqnarray}
where
\begin{eqnarray*}
\Delta_{i,i} & = & - 4 \\
\Delta_{i,j} & = & 1 \ \ \ \ \ \mbox{if i and j are nearest neighbours} \\
\Delta_{i,j} & = & 0 \ \ \ \ \ \mbox{otherwise}
\end{eqnarray*}
The matrix $\Delta$ is the discrete Laplacian. 
Through toppling, neighbouring sites can become unstable, topple themselves, create
new unstable sites, and so on. This {\it avalanche} of topplings proceeds on 
the second, very fast, time scale, so that no new grains of sand are added before
the avalanche is over.
Sand can leave the system when a boundary site topples. Note that (see figure 1)
in the Sierpinski gasket there are only three sites through which sand can
flow out. An avalanche is over when all sites are stable again. We call also
the resulting configuration of height variables stable.
It is not difficult to see that the order in which unstable sites are toppled
doesn't influence the stable configuration which is obtained when the
avalanche is over. This Abelian nature of the sandpile model is crucial in
the further analysis of its properties \cite{Dhar}. 

For further
reference it also necessary to introduce the matrix $G$ which is the inverse
of $\Delta$ and which is therefore the lattice Green function of the
Laplacian.  
It was shown by Dhar that the element $G_{ij}$ of the lattice Green function gives
the expected number of topplings at site $j$ after a grain of sand was dropped
at site $i$.

The criticality of the model manifests itself in power law distributions which
are found for many quantities. For example, let $D(s,N)$ denote the distribution of the number of
distinct sites $s$ toppled in an avalanche. It is found numerically that
$D(s,N)$ obeys a scaling law 
\begin{eqnarray}
D(s,N) \approx s^{-\tau} F(s/N)
\label{2.2}
\end{eqnarray}
Extensive simulations of the sandpile model in two dimensions lead to the estimate $\tau=1.22$
\cite{Manna}.
It is also found numerically that avalanches are compact, i.e. the fractal dimension $D$
of the set of sites toppled in a avalanche 
is equal to that of the lattice on which the model is studied. In particular,
we have for the case of the Sierpinski gasket of figure 1, $D=\ln{3}/\ln{2}$.

The number of stable configurations of the model equals $z_c^N$. However, it has been shown by
Dhar \cite{Dhar} that the number of recurrent configurations in the stationary state is 
exponentially smaller. In fact, there is a one-to-one correspondance between these recurrent
configurations and the number $S_N$ of spanning trees on a lattice ${\cal L}^{\star}$ which 
consists of ${\cal L}$
extended with an extra site, known as the sink, which represents the boundary \cite{DharM}. It is known
that $S_N$ grows in general as $\mu^{N + o(N)}$, 
where $\mu$ is called the {\it connective constant}
of spanning trees \cite{CV}. Spanning trees are described
by the $q$-state Potts model in the limit $q \to 0$ \cite{Wu}. The same model also describes
resistor networks. 

Because of the Abelian nature of the sandpile model, we can perform the
toppling of unstable sites in any desired order. 
A particularly interesting order leads to the concept of {\it waves} \cite{Waves}.
Suppose that an avalanche
starts at a site $i_0$ and that after some topplings $i_0$ becomes unstable
again. One can then keep $i_0$ fixed for the moment and continue with the
toppling of other sites untill all of them (apart from $i_0$) are stable. It is easy to show
that in such a sequence all sites topple at most once. This set of topplings
is called {\it the first wave}. Next, we topple the site $i_0$ for the second time. If after
some topplings it becomes unstable again, we keep it fixed, and topple all
the other unstable sites. This set of topplings constitutes the second wave. We continue in this way
untill we finally reach a stable configuration. We can in this way decompose
any avalanche in a set of waves.
Of course, subsequent waves in an avalanche are highly correlated, but we
can also consider the statistical properties of the set of all waves, irrespective
of the avalanche in which they occur. In this way, we can look at the distribution
$\hat{D}(s_w,N)$ of the number of sites $s_w$ toppled in any wave. Again, it is
found that $\hat{D}(s_w,N)$ has a scaling form
\begin{eqnarray}
\hat{D}(s_w,N) \approx s^{-\tau_w} \hat{F}(s_w/N)
\label{2.3}
\end{eqnarray}
where we have introduced the exponent $\tau_w$ for waves.

The properties of the last wave in a given
avalanche are also of interest. The distribution of the number of sites $s_{lw}$ toppled in this last
wave leads to the introduction of still another exponent, denoted as $\tau_{lw}$.
Because $s_{lw} \leq s$, we arrive at the obvious upper bound
\begin{eqnarray}
\tau \leq \tau_{lw}
\label{3.priem}
\end{eqnarray}
It turns out that the properties of waves are easier to analyse than those of
the avalanches. Without going into details we summarize the following important
results:
\begin{itemize}
\item There is a one-to-one correspondance between waves and two-rooted spanning
trees (one root corresponds with the sink, the other
with the site $i_0$) on ${\cal L}^{\star}$ \cite{Waves}. Furthermore,
the element $G_{ij}$ of the lattice Green function is given by the ratio of the number
of two rooted spanning trees (in which $i$ and $j$ are in the same subtree)
to the number of (one rooted) spanning trees $S_N$. Hence, properties of waves
can be determined from $G$. From the known asymptotic behaviour of the lattice Green function
on an Euclidean lattice, $G_{ij} \sim |i-j|^{2-D}$, one then finds
\begin{eqnarray}
\tau_w = \frac{2D-2}{D}
\end{eqnarray}
\item The exponent $\tau_{lw}$ can be obtained from the probability that
$b$ sites get disconnected from a spanning tree when a randomly chosen bond of the tree is
deleted. In particular, one finds that $\tau_{lw}$ can be expressed in terms of the chemical
dimension $z$ of spanning trees \cite{LastW}. On an Euclidean lattice the precise result is
\begin{eqnarray}
\tau_{lw}= 1 + \frac{2-z}{D}
\label{2.3.b}
\end{eqnarray}
\end{itemize}
From these results, one obtains, in $D=2$ (where $z=5/4$ \cite{DharM})
\begin{eqnarray*}
\tau_w = 1 \ \ \ \ \ \ \ \ \ \ \ \ \ \ \ \ \ \tau_{lw}=11/8
\end{eqnarray*}
 
A final theoretical development which we need to discuss leads to a conjecture on the
exponent $\tau$ \cite{Priez}. Priezzhev {\it et al.} first relate $\tau$ to
another exponent, $\alpha$. To introduce this exponent, it is assumed that the
number of sites toppled in consecutive waves (say the $k$-th  and $k+1$-th wave) decreases.
The change in size,
$\Delta s_w=s_w(k)-s_w(k+1)$ is assumed to occur in a self-similar way, so that
\begin{eqnarray}
\Delta s_w \sim s_w^{\alpha}
\label{2.4}
\end{eqnarray}
One can then immediately show \cite{Priez} that
\begin{eqnarray}
\tau + \alpha = 2
\label{2.5}
\end{eqnarray}
Finally, using the relation between waves and two-rooted spanning trees, and considering 
possible processes by which the size of a wave can decrease, it is found that 
$\alpha=3/4$ and hence $\tau=5/4$. This prediction is in excellent agreement with numerical
estimates, and may very well be exact.

\section{Numerical results}
In order to obtain numerical estimates for the exponents $\tau$, $\tau_{w}$,
$\tau_{lw}$ and
$\alpha$ on the Sierpinski gasket we have performed extensive simulations
of the sandpile model on this fractal structure. In these simulations, we have
considered lattices with $n=3$ up to $n=8$ (the latter contains $N=9843$ sites).
For each of these we have studied more then ($1000\ N$) avalanches. 

In figure 2, we show our results for the distribution of the size of the waves.
As can be seen, log-periodic oscillations are superposed on the power law
decay of the distribution. The behaviour of $D(s,N)$ can therefore be described by complex critical exponents.
This is a well known phenomenon in systems with {\it discrete} scale invariance \cite{DisSI}.
The imaginary part of the critical exponent is in a sense trivial since it can be related to the discrete rescaling
factor of the fractal lattice. From a careful analysis of the data one can hence
obtain finite size estimates of the exponent $\tau_{w}(n)$ as a function of $n$.
These can be extrapolated by making the usual assumption
\begin{eqnarray}
\tau_{w}(n) = \tau_w + \frac{C}{\ln n}
\label{2.6}
\end{eqnarray}
In figure 3 we show this extrapolation. It leads to the estimate
\begin{eqnarray}
\tau_w = 0.47 \pm 0.05
\label{2.7}
\end{eqnarray}
It is difficult to make a precise determination of the error in this estimate. The
error we give is therefore more indicative.

Figure 4 shows our results for the distribution of the size of the last wave.
From an examination of these data, we find
\begin{eqnarray}
\tau_{lw} = 1.66 \pm 0.05
\label{2.8}
\end{eqnarray}

Finally, our data for the distribution of avalanche sizes, shown in figure 5, yield the exponent estimate
\begin{eqnarray}
\tau = 1.46 \pm 0.05
\label{2.9}
\end{eqnarray}
in agreement with the result found by Kutnjak-Urbanc {\it et al.} \cite{Zap}, who obtained
$\tau=1.51 \pm 0.04$.

We have also investigated the behaviour of $\Delta s_w$ in order to determine
the exponent $\alpha$. To obtain this exponent, we have to consider only those
cases in which $\Delta s_w > 0$ (in fact, there are many cases for
which the opposite inequality holds).
In figure 6 we show our data (after integration over bins).
As can be seen from the figure, it is not easy to determine a clear value
of $\alpha$. Our best estimate is
\begin{eqnarray}
\alpha= -0.15    \pm 0.1   
\label{2.10}
\end{eqnarray}
In the next section, we will give a lower bound for $\alpha$. Our estimate (\ref{2.10})
is very close to that bound. It is however difficult to imagine how $\alpha$ can be
negative. We therefore turned to an
alternative way to determine $\alpha$. This is done by investigating how the average number $n_w$ of
waves in an avalanche scales with the size of the avalanche. When the size
of consecutive waves decreases, the size of the first wave is proportional
to the size of the whole avalanche. From (\ref{2.4}) we then obtain
\begin{eqnarray}
n_w \sim s^{1 - \alpha}
\end{eqnarray}
In figure 7 we show data for this quantity for $n=6$ and $n=8$.

 We find
{\it very clear evidence for a linear relation between $n_w$ and $s$}.
This relation breaks down only for the largest avalanches, which are those that span the lattice. But, as
pointed out by Kutnjac-Urbanc {\it et al.} \cite{Zap}, these avalanches are somewhat peculiar.
They occur here because only through such avalanches is it possible for
sandgrains to leave the system.

From the linear relation we find
\begin{eqnarray}
\alpha = 0.0 \pm 0.01
\label{2.ex}
\end{eqnarray}
These data lead us to conjecture that $\alpha=0$ exactly. As we will see in the
next section, this result will be very useful in determining the
exponent $\tau$.

\section{Exact and conjectured results}
We now present some exact results for the connective constant $\mu$ for
spanning trees ( and hence for the number of recurrent states) and for the
exponents $\tau_{w}$ and $\tau_{lw}$. We make also a conjecture for $\tau$.

As discussed in section 2, there exists a close connection between the theory
of the sandpile model and the ($q \to 0$)--limit of the Potts model. This connection
should allow the determination of several properties of the sandpile from 
a study of that model on the Sierpinski gasket. For example, the connective constant
of the spanning trees can be obtained from the free energy of the 
Potts model when $q \to 0$. This free energy in turn can most easily be obtained
by performing a simple but exact real space RG calculation for the Potts model on 
the Sierpinski gasket. 
Consider therefore the $q$-state Potts model for which we will denote the
spin variable at vertex $i$ by $\sigma_i$. On a Sierpinski gasket, these
spins can also be labeled according to the `generation' in which
they appear during the iterative construction of the fractal. 
The reduced Hamiltonian $H(K)$ of the Potts model is given by  
\begin{eqnarray}
H(K) = K \sum_{\langle i,j \rangle} \delta_{\sigma_i,\sigma_j}
\label{2.11}
\end{eqnarray}
A renormalisation group for $H(K)$ 
on the Sierpinski gasket
 can be obtained
by performing a partial trace, denoted by $\mbox{Tr}^{\star}$ over the spin variables which have the
same `generation label':
\begin{eqnarray}
e^{\ds H(K') + g(K)} = \mbox{Tr}^{\star} e^{\ds H(K)}
\label{2.12}
\end{eqnarray}
The free energy of the model can then be obtained by a summation of $g(K)$ along
the renormalisation group trajectory.
We have performed such a calculation. After sending $q \to 0$ the RG flow is
remarkably simple
\begin{eqnarray}
\ & \ & K' = \frac{3}{5} K \label{2.12.b}\\
\ & \ & e^{\ds g(K)} = 50 K^3 q^{3/2}
\end{eqnarray}
from which we obtain the connective constant for spanning trees after a straightforward calculation.
The result is
\begin{eqnarray}
\mu = 50^{1/3} \sqrt{3/5} \approx 2.854
\label{2.14}
\end{eqnarray}

As mentionned in section 2, the exponent $\tau_{lw}$ can be obtained from the 
probability $P_s(b)$ that upon deleting a bond at random from a spanning tree, 
a subtree of size $b$ gets disconnected. On an Euclidean
lattice this relation leads to the result (\ref{2.3.b}).
On a fractal lattice, the relation has to be modified. The correct expression
now also involves the fractal dimension
$D_w$ of a random walk on the fractal. This was shown in reference \cite{DharDhar},
where it was found that
\begin{eqnarray*}
P_s(b) \sim b^{-D_w/z}
\end{eqnarray*}
Then, repeating the arguments in \cite{LastW} we obtain
\begin{eqnarray}
\tau_{lw}= 1 + \frac{D_w-z}{D}
\label{2.3.c}
\end{eqnarray}
This is a quite general result for $\tau_{lw}$ which should for all cases (fractal
and Euclidean).
 
For spanning trees on the Sierpinski gasket, $z$ was recently determined by Dhar and Dhar
in a study of loop-erased random walks \cite{DharDhar}. Their result is 
\begin{eqnarray}
z= \frac{\ln{[(20 + \sqrt{205})/15)]}}{\ln 2}
\label{2.15}
\end{eqnarray}
[Before proceeding, we would like to point out a general recipe for determining $z$,
using again an RG approach.
 For a tree, the chemical dimension equals the red bond
dimension $D_R$. The red bonds of a connected graph are those edges which when cut, disconnect
the graph. Coniglio \cite{Con}, and independently Saleur and Duplantier \cite{DS}, pointed out that
the red bond dimension for the critical Potts model can be obtained by adding
a term
\begin{eqnarray*}
M \sum_{i,j} \delta_{\sigma_i,1}\delta_{\sigma_j,1}
\end{eqnarray*}
to the Potts hamiltonian (\ref{2.11}). The red bond dimension can then be obtained from
the linearised RG equation for $M$ at the critical Potts fixed point.
Hence, it is possible to get $z$ for spanning trees by performing an RG calculation
(either exact or approximate)  on an extended Potts model. We will present results
from such an approach in a future publication.]

Inserting (\ref{2.15}) in (\ref{2.3.c}) and using the value of $D_w$ for
the Sierpinski gasket ($D_w=\ln{5}/\ln{2}$, \cite{DSier}) we get
\begin{eqnarray}
\tau_{lw} = 1 - \frac{\ln{[(20 + \sqrt{205})/75]}}{\ln 3} \approx 1.712
\label{2.16}
\end{eqnarray}
This prediction is in good agreement with the numerical value found in section 3.

Next, we turn to the exponent $\tau_w$. This should be determined from the behaviour
of the Green function. We will assume that the Green function $G_{ij}$ decays for large $r=|i-j|$
as 
\begin{eqnarray}
G_{ij} \sim r^{-2 x}
\label{2.17}
\end{eqnarray}
The Green function gives the expected number of topplings at site $j$ when a grain of
sand has been dropped at site $i$. Hence it consists of all waves that cover the
distance between site $i$ and $j$. Since waves are compact, we obtain
\begin{eqnarray*}
G_{ij} \sim \int_{r^D}^{\infty} \hat{D}(y) dy
\end{eqnarray*}
where we have neglected finite size effects.
From (\ref{2.17}) we then get a general expression for the exponent $\tau_w$
\begin{eqnarray}
\tau_w = 1 + \frac{2x}{D}
\label{2.18}
\end{eqnarray}
To determine $\tau_w$ for the particular case of the Sierpinski gasket,
we only need to determine the value of $x$. This exponent was determined in
early studies on the behaviour of the resistance on fractals with the
result \cite{ResS}
\begin{eqnarray*}
2x = \frac{\ln{(5/3)}}{\ln{2}}
\end{eqnarray*}
This leads to the prediction
\begin{eqnarray}
\tau_w = \frac{\ln{(9/5)}}{\ln{3}} \approx 0.535
\label{2.19}
\end{eqnarray}
This value is again in agreement with the one determined numerically.
In principle, the exponent $x$ can also be obtained from the $q \to 0$ limit of
the Potts model. In fact, it can be obtained from the renormalisation group
equation for $K$, see (\ref{2.12.b}).

Finally, we turn to the exponents $\tau$ and $\alpha$.
It is easy to generalise the result (\ref{2.5}) which was obtained in $D=2$. 
The size $s_w$ of any wave must be less then or equal to the size $s$ of
the avalanche in which it occurs. Moreover, the number of waves whose size lies
between $s_w$ and $s_w + ds_w$ is according to (\ref{2.4}) proportional to
$s_w^{-\alpha}$. Combining these two results we obtain for the size
distribution of waves
\begin{eqnarray*}
\hat{D}(s_w) \sim s_w^{-\alpha-\tau-1}
\end{eqnarray*}
from which we immediately obtain the appropriate generalisation of (\ref{2.5})
\begin{eqnarray}
\alpha + \tau = 1 + \tau_w
\label{2.20}
\end{eqnarray}
Our numerical results (\ref{2.7}), (\ref{2.9}) and (\ref{2.ex}) are nicely
consistent with this picture. From the upper bound (\ref{3.priem}), we obtain a
bound on $\alpha$
\begin{eqnarray*}
\alpha > 1 + \tau_w - \tau_{lw} \approx -.177
\end{eqnarray*}
Our first estimate of $\alpha$, given in (\ref{2.10}), is just consistent with this bound.

Most interestingly our conjecture that $\alpha=0$ leads to a conjecture
for $\tau$
\begin{eqnarray}
\tau = 1+ \tau_w = \frac{\ln{(27/5)}}{\ln{(3)}} \approx 1.535
\label{2.21}
\end{eqnarray}
which is in excellent agreement with our numerical results.

We have also tried to extend the argument of Priezzhev and his coworkers for
the exponent $\alpha$ to the fractal case. As far as we could see, such an argument
always leads to a value of the exponent $\alpha$ which is of order one, and
which therefore is completely in disagreement with the numerical result.
It therefore seems that the way in which consecutive waves are related is
different for the fractal case than for the Euclidean one.

\section{Discussion and conclusions}
In this paper, we have presented extensive simulations for waves and avalanches
on the Sierpinski gasket. One of the most surprising results is the fact
that the exponent $\alpha$ is very close to zero. This implies that
the way in which waves become smaller is almost independent of their size.
This is most probably a direct consequence of the fractal nature of the
Sierpinski lattice.

We have also extended the existing theory for waves to the present case. This
has led to exact expressions for the exponents $\tau_{lw}$ and $\tau_w$ which
are close to the numerically determined values.

Most interestingly, assuming that $\alpha$ is zero, we are led to the conjecture
(\ref{2.21}) for the exponent $\tau$. Also this prediction is in good agreement
with our simulation results.

In this way, there is now a second situation (after the two-dimensional case) for
which wave exponents are determined exactly and for which a conjecture for
the $\tau$-exponent exists.

The result $\alpha=0$ indicates, in our opinion, that there is good hope that
the sandpile model on a Sierpinski gasket may be solved exactly. We are currently
using real space RG approaches which have been introduced for sandpile models \cite{Piet,Wies},
to see whether they may be worked out exactly on this lattice.

\ \\
{\bf Acknowledgement} One of us (CV) would like to thank D. Dhar and V. Priezzhev
for useful discussions on the subject of this paper. We also thank the Inter University
Attraction Poles for financial support.

\begin{figure}

  \centerline{ \psfig{figure=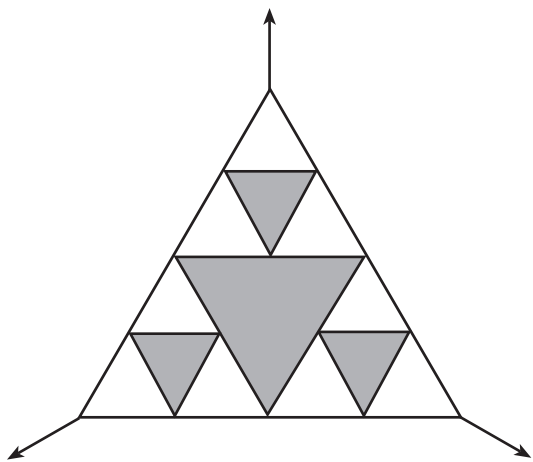,width=7cm}}

\caption{The simple Sierpinski gasket with $n=2$. Sand can only leave the lattice
through the three sites indicated by an arrow.}

\label{figure 1}
\end{figure}

\begin{figure}
  \centerline{ \psfig{figure=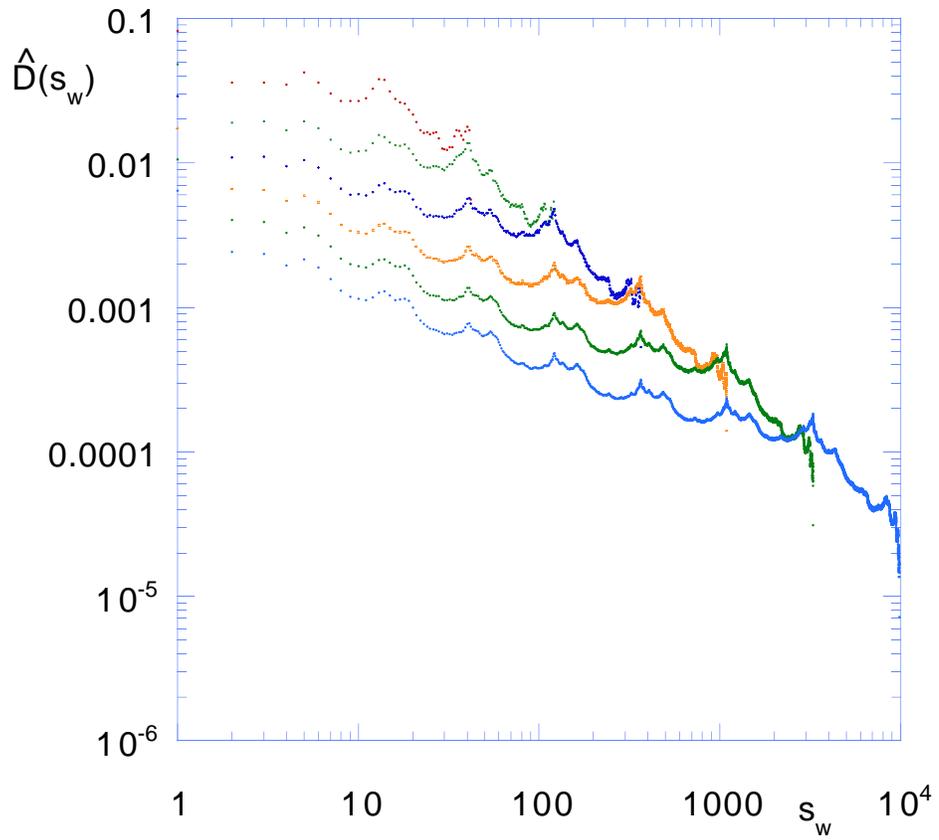,width=15cm}}

\caption{Distribution of wave sizes. The data give numerical results for
$n=3$ to $n=8$ (top to bottom).}

\label{figure 2}
\end{figure}

\begin{figure}
  \centerline{ \psfig{figure=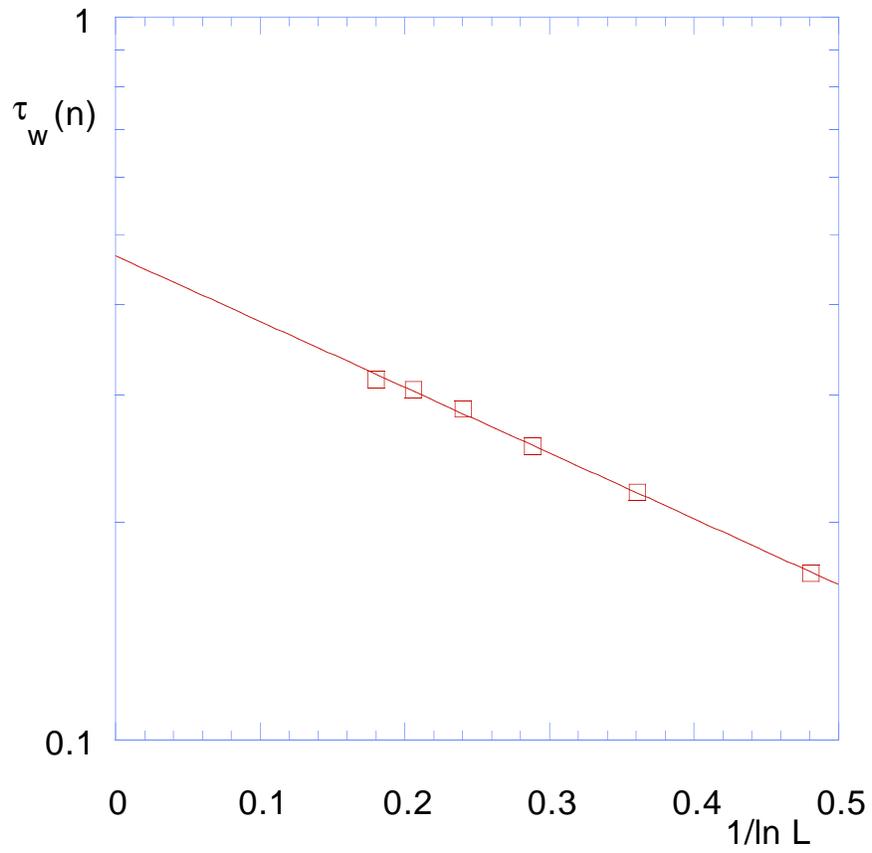,width=15cm}}
\caption{Extrapolation of $\tau_w(n)$ ($L=2^n$).}

\label{figure 3}
\end{figure}

\begin{figure}
  \centerline{ \psfig{figure=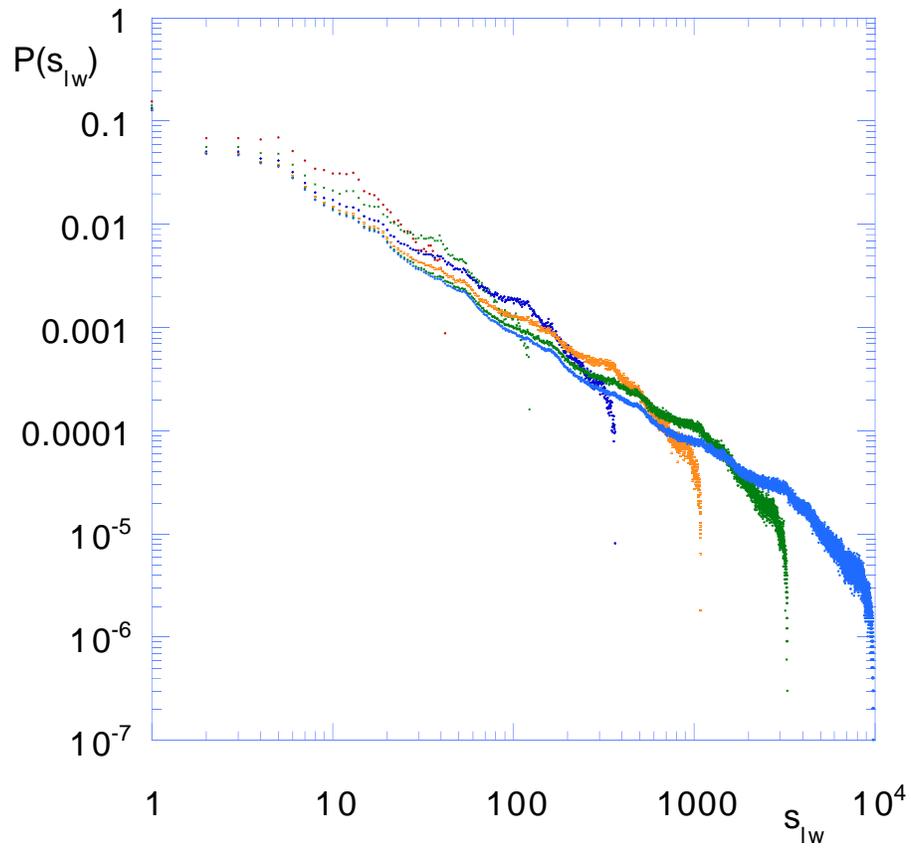,width=15cm}}
\caption{Distribution of last wave sizes. Data are given for $n=3$ to $n=8$.}

\label{figure 4}
\end{figure}

\begin{figure}
  \centerline{ \psfig{figure=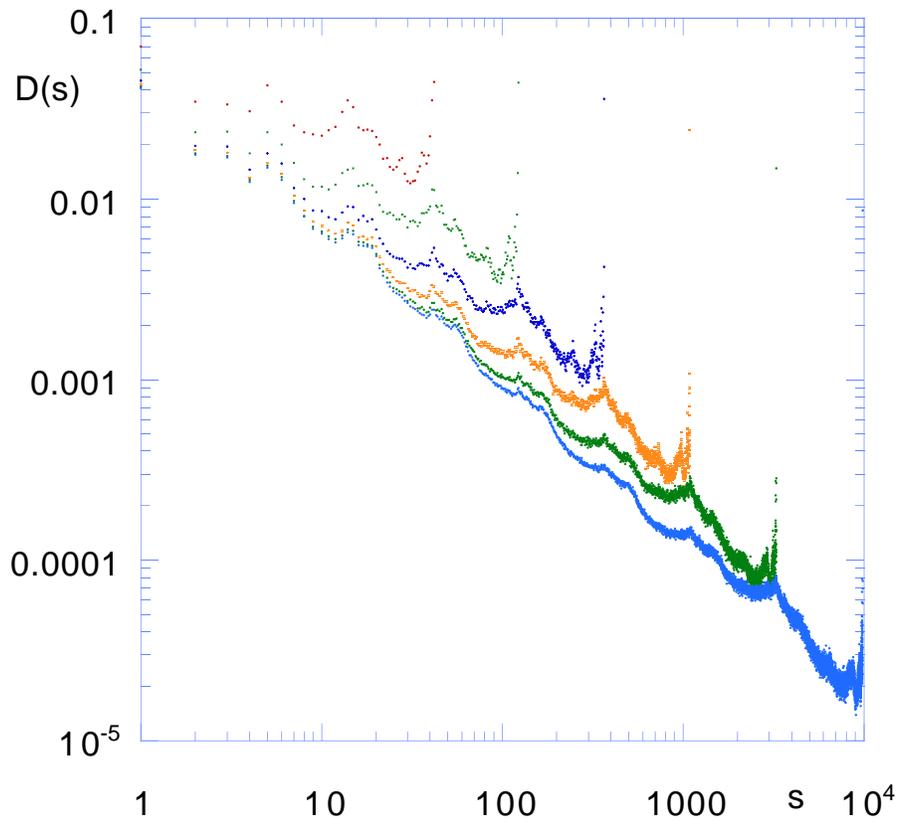,width=15cm}}
\caption{Distribution of avalanche sizes. Data are given for $n=3$ to $n=8$.}

\label{figure 5}
\end{figure}

\begin{figure}
  \centerline{ \psfig{figure=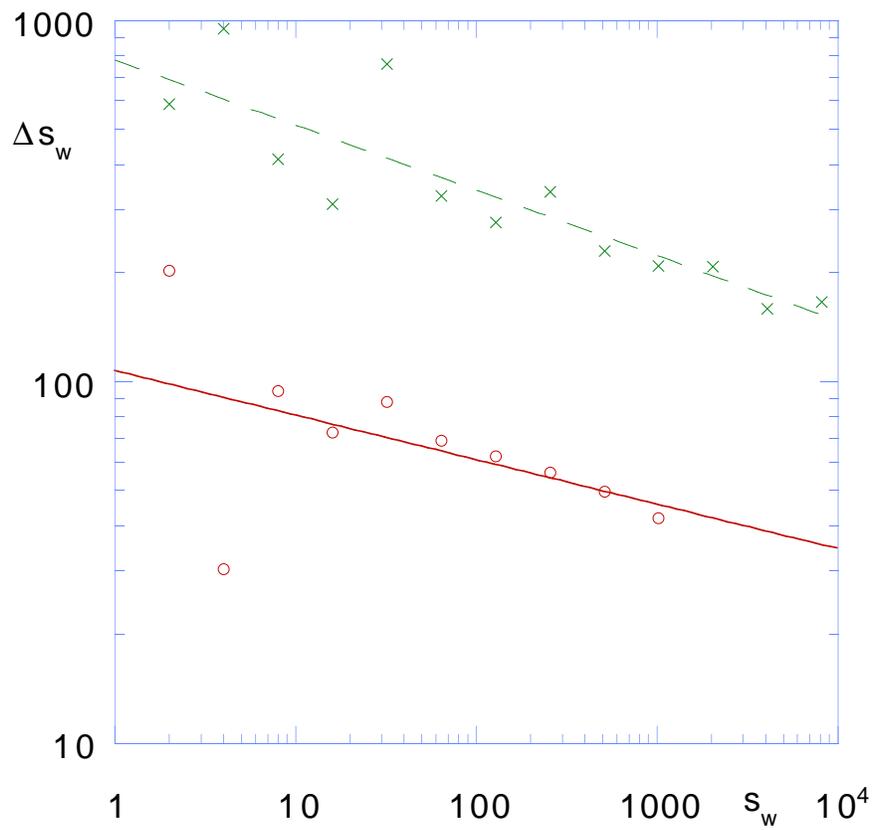,width=15cm}}
\caption{Decrease in size (averaged over bins) of consecutive waves, versus wave size,
for $n=6$ and $n=8$ (upper dataset).}

\label{figure 6}
\end{figure}

\begin{figure}
  \centerline{ \psfig{figure=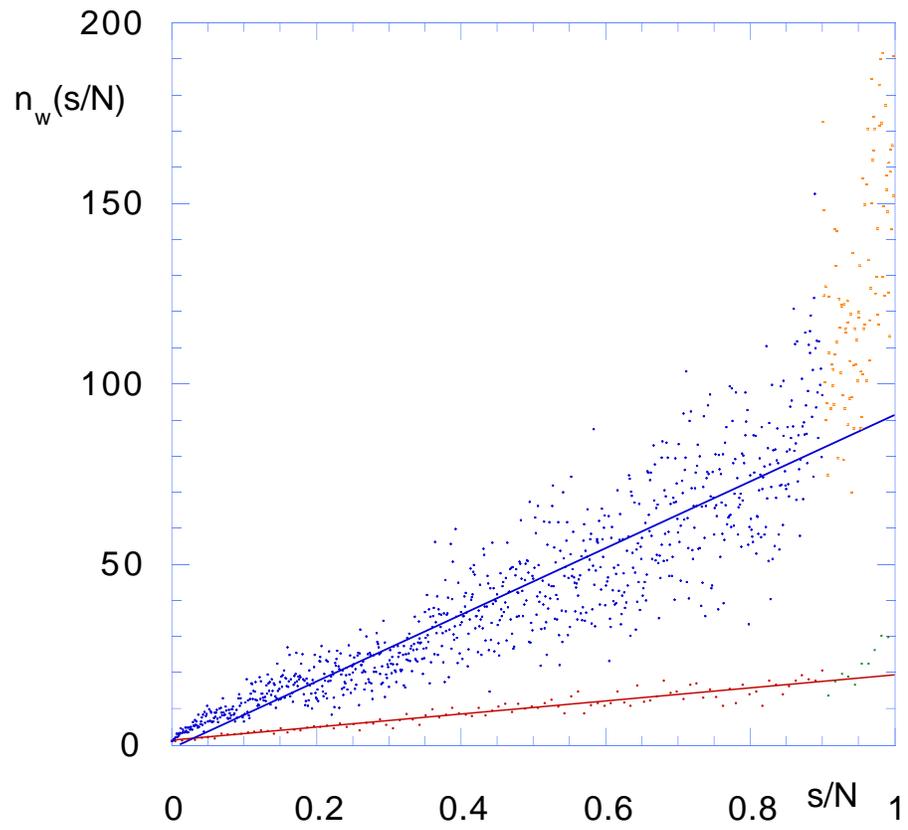,width=15cm}}
\caption{Number of waves in an avalanche versus size of the wave, for
$n=6$ and $n=8$ (upper dataset).}

\label{figure 7}
\end{figure}

\end{document}